\documentclass[journal]{IEEEtran}
\usepackage{balance}

\usepackage{cite}
\usepackage{subfigure}

\ifCLASSINFOpdf
  \usepackage{graphicx}
\usepackage{amsmath}

\usepackage{stfloats}
\usepackage{amssymb}

\newcommand{\smg}{\widehat{S}_{{mG}}}
\usepackage{fancyhdr,mwe,lipsum,array}
\newcolumntype{P}[1]{>{\centering\arraybackslash}p{#1}}
\begin{document}
\title{Multitapered Modified Group Delay Function for Representation of Speech Signals}

\author{Narendra~K.~C.,~\IEEEmembership{Student Member,~IEEE,}
        R.~Kumaraswamy,~\IEEEmembership{Senior~Member,~IEEE,}
        and~Sanjeev~Gurugopinath,~\IEEEmembership{Member,~IEEE}
\thanks{Narendra~K.~C.~and S.~Gurugopinath are with the Department of Electronics and Communication Engineering, PES University, Bengaluru 560085, India. E-mails: \{narendrakc, sanjeevg\}@pes.edu}
\thanks{R.~Kumaraswamy is with Department of Electronics and Communication Engineering, Siddaganga Institute of Technology, Tumkur 572103, India. E-mail: rks@ieee.org}
\thanks{Manuscript received April 19, 2005; revised August 26, 2015.}}

%


\maketitle

\begin{abstract}
In this paper,  a novel multitaper modified group delay function-based representation for speech signals is proposed. With different set of phoneme-based experiments, it is shown that the proposed method performs better that an existing multitaper magnitude (MT-MAG) estimation technique, in terms of variance and MSE, both in spectral- and cepstral-domains. In particular, the performance of MT-MOGDF is found to be the best with the Thomson tapers. Additionally, the utility of the MT-MOGDF technique is highlighted in a speaker recognition experimental setup, where an improvement of around $20\%$ compared to the next-best technique is obtained. Moreover, the computational requirements of the proposed technique is comparable to that of MT-MAG.  The proposed feature can be used in for many speech-related applications; in particular, it is best suited among those that require information of speaker and speech.
\end{abstract}


\IEEEpeerreviewmaketitle

\section{Introduction}

Representation of a speech signal is challenging, as it contains manifold  information about speaker, speech, language, accent, emotion, etc. . While the speaker information is normally represented by the fine harmonic structure in the spectrum, the speech information is embedded in the envelope of the spec. Spectral-domain representations are widely employed in many applications, and hence obtaining a good estimate of speech spectrum is a key requirement. A simple procedure is to compute the DFT of the autocorrelation sequence of a given speech signal, which is known as the periodogram estimate. However, the periodogram estimate has a large variance \cite{yegnanarayana1992significance}, and a better estimate can be obtained by multiplying the frame of the speech signal with a window function that is tapered toward the ends of the signal frame. The variance of the estimated spectrum can be further reduced by averaging multiple realizations of the same random process \cite{welch1967use}, which in the case of speech corresponds to same utterances spoken by either a single or different speakers. Use of multiple orthonormal windows \cite{thomson1982spectrum}, instead of a single window as proposed in the former method, yields lower variance by combining the advantages of windowing and averaging. However, this also reduces the spectral resolution.  We term this technique as multitaper magnitude (MT-MAG) based estimation, whose analysis and performance benefits are extensively discussed in \cite{kinnunen2012low}.

Alternatively, the phase spectrum of a speech signal contains information about the resonances of the vocal tract in its transitions, and is known to provide a representation which is similar to the magnitude spectrum \cite{yegnanarayana1984significance}. Towards this end, the group delay function (GDF) \cite{yegnanarayana1984significance} was defined, which obtains the transitions in the phase spectrum by computing its negative derivative. It is known that the peaks and valleys in the GDF of speech corresponds to the resonances and anti-resonances of the vocal tract, respectively. Additionally, the GDF accurately captures the Formant information in the speech signal, since the phase spectrum is immune to additive degradations on the speech signal \cite{murthy2011group}. However, the GDF suffers from spurious peaks that are caused due to the zeros located near the unit circle of the $z$-plane. To alleviate this, a two-parameter modified group delay function (MOGDF) was proposed in [], which used a cepstrally smoothed version of the spectrum is employed to reduce the spuriousness in the GDF.

In this paper, a novel speech representation is proposed by processing the phase spectrum -- in terms of the MOGDF -- of speech signals multiplied with a set of orthonormal tapers. This new representation, termed as the multitaper MOGDF (MT-MOGDF), has the collective advantage of MOGDF and performance benefits obtained by employing multiple tapers. Through experimental results, we show that the proposed technique has lower variance and lower mean squared error (MSE) as compared to the MT-MAG technique, in both spectral- and cepstral-domains. Also, the high resolution property of MOGDF is retained in this representation, which is useful in several applications. As a case study, an application of speaker recognition with varying number of speakers is considered, and is shown that MT-MOGDF performs the best as compared to all the other techniques listed above, owing to its low variance advantage.



 The remainder of the paper is organized as follows. The existing spectral estimation techniques and the proposed multitaper modified group delay representation are given in Sec.~\ref{sec:mtmogdf}. A detailed study on the performance of the proposed technique in comparison to MT-MAG with bias, variance and mean squared error measures for different types and number of tapers is discussed in Sec.~\ref{sec:experiments}. Performance improvements due to MT-MOGDF as compared to the other techniques for a speaker recognition application are highlighted in Sec.~\ref{sec:spkrec}. 

 

\section{Multitaper Modified Group Delay Function Representation}

\label{sec:mtmogdf}
  
 The  GDF provides a meaningful representation of the vocal tract information by processing phase spectrum of the speech signal\cite{murthy2011group}. 
Consider the spectrum of a speech signal $s(t)$ given by 
\begin{equation}
\label{eqn:FT}
S(\omega)=|S(\omega)| e^{\theta(\omega)}, ~~\omega \in \mathbb{R},
\end{equation}

where $|S(\omega)| $ and $\theta(\omega)$ are  the magnitude and  phase spectra of the speech signal, respectively. The GDF  is defined as
\begin{equation}
\label{eqn:GDF}
\tau(\omega)=-\frac{d\theta(\omega)}{d\omega}.
\end{equation}
 The GDF exhibits near-squared magnitude behaviour at the resonances, but exhibits spurious spikes if the zeros are located near unit circle in the $z$-plane. To minimize the effect of these peaks, a modified group delay function (MOGDF) was proposed, which alleviates the spuriousness by scaling down the power spectrum \cite{hegde2007significance}, and by radially moving the zeros towards the origin. The MOGDF is defined as
\begin{equation}
\label{eqn:MOGDF}
\tau_m(\omega)=\frac{\tau(\omega)}{|\tau(\omega)|}\left(|\tau(\omega)|\right)^\alpha, ~~0< \alpha \leq 1,
\end{equation}
where
\begin{equation}
\label{eqn:tau}
\tau(\omega)=\frac{X_R(\omega)Y_R(\omega)+X_I(\omega)Y_I(\omega)}{|\xi(\omega)|^{2\gamma}},~~0< \gamma \leq 1.
\end{equation}

Here, $X(\omega)$ is the Fourier spectrum of a speech signal frame $x(n)$, and $Y(\omega)$ is the Fourier spectrum of $nx(n)$. The subscripts $R$ and $I $ denote the real and imaginary parts of the spectrum. $\xi(\omega)$ is the cepstrally smoothed spectrum of $S(\omega)$. Parameters 
$\alpha$ and $\gamma$ are  tuned to reduce the spurious spikes in the group delay spectrum.

 A simple estimate of the spectrum is obtained by computing  the DFT of the autocorrelation sequence of a given speech signal, known as the periodogram estimate. However, the periodogram estimate has a large variance \cite{yegnanarayana1992significance}, which can be reduced either by multiplying the time-domain speech signal by a tapered window, or by averaging over multiple realizations. The multitaper (MT) technique  \cite{kinnunen2012low} combines both these advantages. An MT-based periodogram spectral estimate was proposed in \cite{kinnunen2012low}, which was shown to exhibit smoothness with lower variance. Termed as MT-based magnitude spectrum (MT-MAG), this estimate is given by \cite{kinnunen2012low}
 \begin{equation}
 \label{eqn:mtmag}
 \widehat{S}(\omega)=\sum_{j=1}^{N}\lambda(j)\left|\sum_{t=o}^{M-1}w_j(t)x(t)e^{-i2\pi tf/N_f}\right|^2,
 \end{equation}
 where $x(t)$ is the time-domain speech signal, $w_j =[w_j(0),w_j(1),\dots, w_j(M-1)]^T$ is a set of $N$ orthogonal tapers, and $\lambda(j)$ is a weight assigned to each of the tapers, $j=1,2,3,\dots,N$.  These tapers are chosen such that the errors in estimating the sub-spectra are nearly uncorrelated in frequency-domain. It is known that the variance of the multitaper magnitude (MT-MAG) spectrum follows
 \begin{equation}
 V[\hat{S}(\omega)]\approx \frac{1}{N}|S(\omega)|^2.
 \label{eqn:mtvar}
 \end{equation}

In this paper, a novel multitaper MOGDF (MT-MOGDF) function for spectrum estimation in speech signals is proposed. The performance is analyzed in terms of a bias-variance study. The MT-MOGDF estimate of $\widehat{S}(\omega)$ is defined as
\begin{equation}
\smg(\omega)=\sum_{j=1}^N \lambda(j)\tau_{jm}(\omega)
\label{eqn:mtmogdf}
\end{equation}
where $\tau_{jm}(\omega)$ is the MOGDF function for the speech frame as given in (\ref{eqn:MOGDF}) is  multiplied by the $j^{th}$ taper and $\lambda(j)$  is the associated weight of the taper.  Further, the choice of the tapers based on the number and shape is elaborated in \cite{kinnunen2012low}. To analyze the performance of the MT-MOGDF,  the bias, variance and mean squared error (MSE) of the log spectrum are defined as
\begin{align}
&B[\hat{S}_{mG}(\omega)]= \mathbb{E}[\log(\hat{S}_{mG}(\omega))]-\log(S_T(\omega)),\\
&V[\hat{S}_{mG}(\omega)]= \mathbb{E}[\log(\hat{S}_{mG}(\omega)^2)]-\mathbb{E}[\log(\hat{S}_{mG}(\omega))]^2,\\
&MSE(\hat{S}_{mG}(\omega))= B[\hat{S}_{mG}(\omega)]^2+V[\hat{S}_{mG}(\omega)].
\label{eqn:perf}
\end{align}
Speech signals are modeled as concatenation of basic units in sound perception called phonemes, which are further divided into vowels and non-vowels. The vowels form a significant part of the speech signal as the energy concentration is more in these regions. Spectrum in vowel-like regions of the speech signal has peaks centred around the resonances of the vocal tract. Also, there would be no constrictions along the vocal tract when the vowels are produced. Hence, the formant locations are significant for spectrum estimation. The proposed MT-MOGDF exploits this key requirement. Among the non-vowels, an extreme type are the sounds of fricatives. These sounds have a flattened spectrum and varies largely across speakers. Therefore, energy-based representations may not be sufficient to model these sounds.It is  shown that MT-MOGDF provides a better representation even in this case, with an appropriate choice of taper.

Most applications in speech processing use the features in cepstral-domain as a better representation. Moreover, the variance and bias improvements in spectral-domain are intuitively reflected in the cepstral domain. Towards this end,  a detailed performance study is considered in the cepstral-domain, as elaborated in the next section. The cepstrum is computed as 
\begin{equation}
\widehat{C}=\textbf{DCT}\{ \log(\smg(\omega))\}.
\label{eqn:ceps}
\end{equation}
$\textbf{DCT}$ is used for compactness in the representation \cite{parthasarathi2011robustness}.


\section{Experimental results and Discussion } 
\label{sec:experiments}
The phoneme instances are obtained from DR7 directory of TIMIT database []. 250 instances of the vowel /a/, and fricative /s/  are  collected from different speakers uniformly randomly. Frames of 10 ms with an overlap of 50\% is considered. This set of frames serves as a set of multiple realizations of the same underlying random process. These are used to obtain the sample mean of the speech signal, whose spectrum is assumed to be the true mean. A total of $N_s=11204$ for /a/, and $N_s = 15064$ for /s/ instances are extracted. 

\begin{figure*}
 
\centering

\includegraphics[width=16cm, height=10cm]{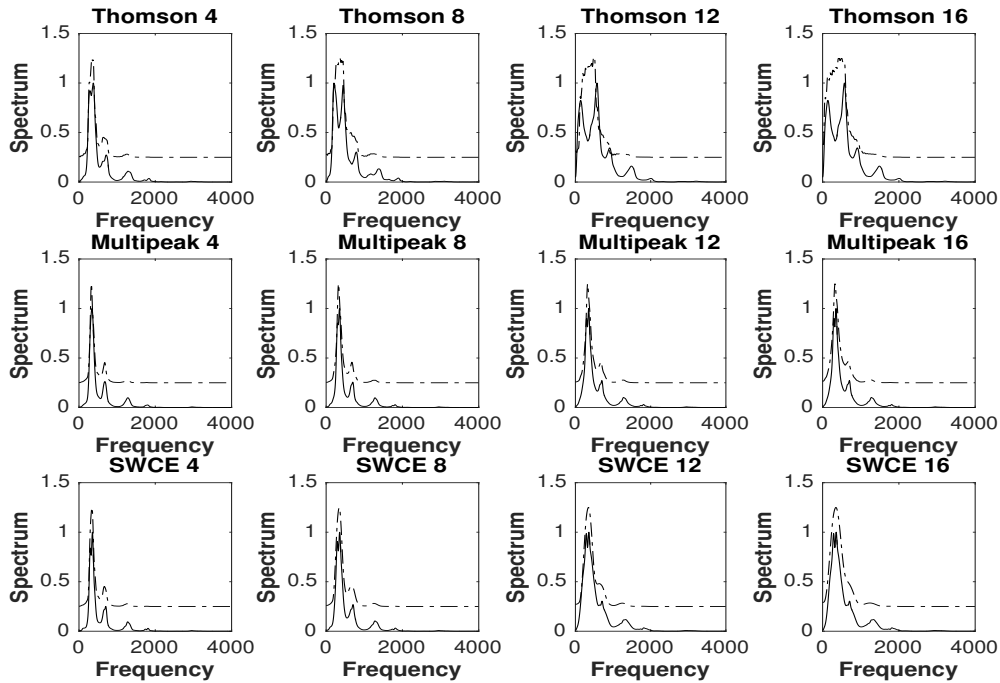}
\vspace{-0.5cm}
\caption{Estimated spectrum for a speech frame in a vowel-like region. The MT-MOGDF simultaneously restores smoothness in spectrum like MT-MAG, and preserves the Formant location information, unlike MT-MAG.}
\label{fig:spec}
\end{figure*}

The MT-MOGDF spectrum is computed as given in (\ref{eqn:mtmogdf}). Comparison of MT-MAG and MT-MOGDF estimates for the above setup with different types of tapers, namely, Thomson, SWCE and multipeak tapers are shown in Fig.~\ref{fig:spec}. The MT-MAG spectrum is boosted by 5 dB for visual clarity. It can be observed that MT-MOGDF and MT-MAG are similar in tracking the envelope of the speech signal. However, the number of formants obtained in MT-MOGDF is better which serves as an added advantage over MT-MAG. This is expected, owing to the high resolution property of the MOGDF function \cite{yegnanarayana1984significance}. Note that the comparison with the periodogram estimate is omitted since MT-MAG is known to outperform the periodogram-based estimate in terms of bias and variance \cite{kinnunen2012low}. Note that the spectrum obtained by MT-MAG is smoother, and as $N$ increases, MT-MOGDF also gives a smoother spectrum. Additionally, information on formant locations is lost in MT-MAG as $N$ increases, while it is retained in MT-MOGDF.

\begin{figure}
 
\centering
\includegraphics[width=8cm, height=5.5cm ]{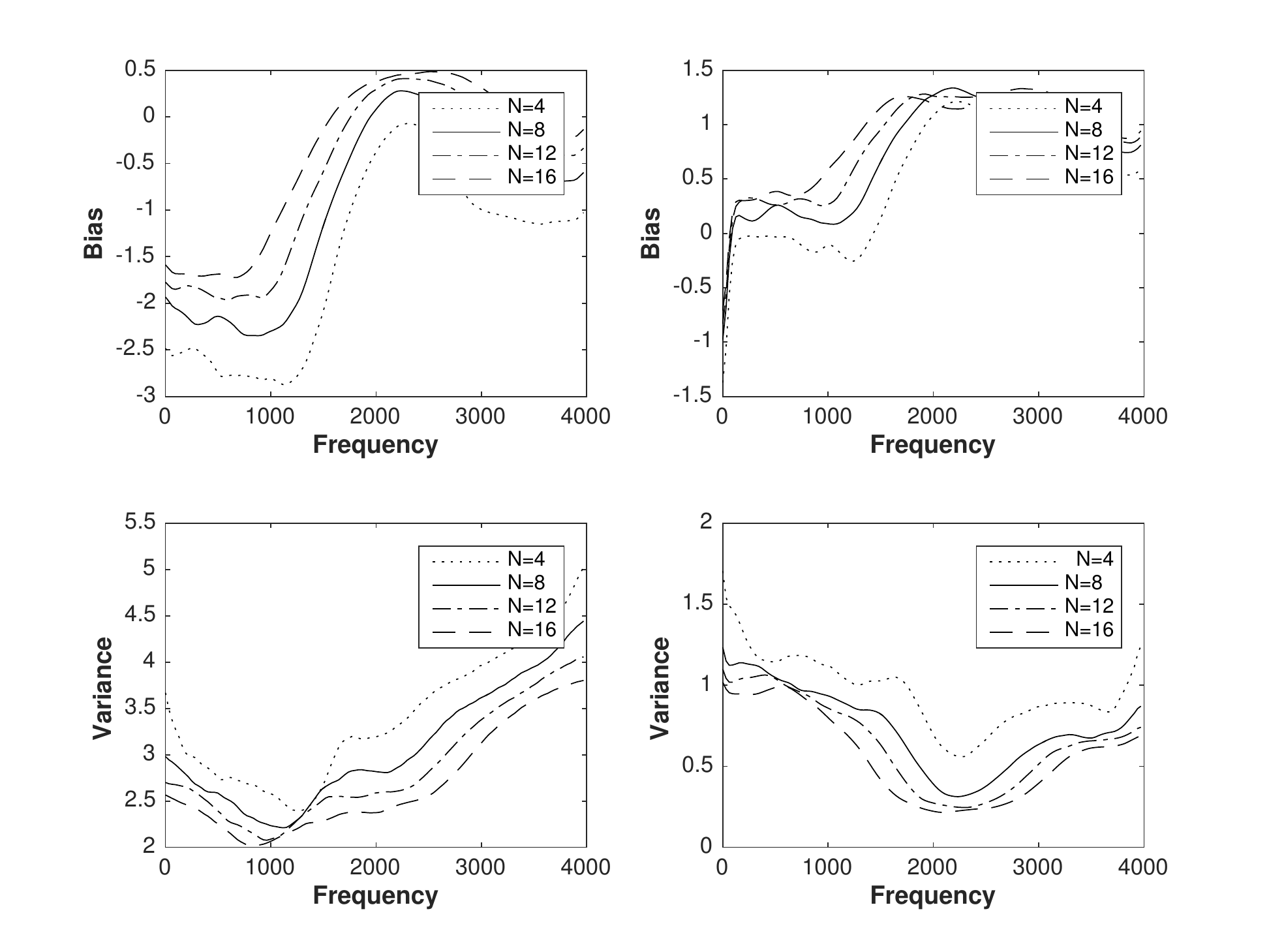}
\vspace{-0.4cm}
\caption{Bias and variance of the spectrum for Thomson taper with MT-MAG (left), and MT-MOGDF (right) for different number of tapers for the phoneme /a/ .  MT-MOGDF best-fit representation leads to a larger bias, but lesser variance in comparison to MT-MAG best-fit representation.}
\label{fig:bstcase}
\end{figure}

\begin{figure}
\centering
\includegraphics[width= 8cm, height=5.5cm]{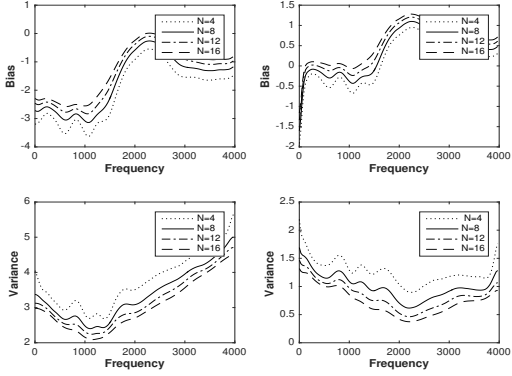}
\vspace{-0.4cm}
\caption{Bias and variance of the spectrum for multipeak taper with MT-MAG (left), and  MT-MOGDF (right) for different number of tapers for the phoneme /a/ .  Even in this case, the larger bias - lesser variance trend of MT-MOGDF representation in comparison to MT-MAG representation is observed.}
\label{fig:wrstcase}

\end{figure}

Next, the performances are studied of MT-MOGDF and MT-MAG in terms of bias and variance measures.  The performance comparison of these two methods is considered with Thomson and multipeak tapers in Figs.~\ref{fig:bstcase} and \ref{fig:wrstcase}, respectively. In both cases, it is observed that MT-MOGDF gives a lesser variance, albeit with a larger bias. An important key feature of the multitaper technique, i.e., reduction in variance with $N$ as governed by \eqref{eqn:mtvar}, is retained even in the case of MT-MOGDF. In other words, the trend dictated by using the multitaper \cite{hansson1997multiple} technique holds good irrespective of whether magnitude or phase is employed for estimation. A trend of lower variance for a  larger band frequency  can be observed in MT-MOGDF than the MT-MAG technique. This is important as the Formats located in frequencies that are greater than 1 kHz should be captured. Usage of a multipeak taper almost uniformly reduces variance over all frequencies. The MT-MOGDF technique works especially well with the Thomson taper, because of its ability to retain the Formant structure. This bias-variance tradeoff between the two techniques becomes clear with the notion of MSE, which will be explored next. Note that the MSE depends directly on the variance and square of the bias term, and hence the sign of the bias is also important.

\begin{figure}
\centering
\includegraphics[width=9cm, height=6cm]{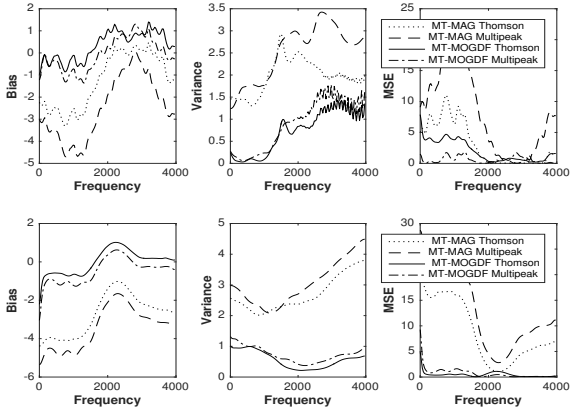}
\vspace{-0.4cm}
\caption{Bias, variance and MSE of the spectrum for the phonemes /a/ (top row) and /s/ (bottom row). The MT-MOGDF provides a lower MSE in comparison to MT-MAG, for different tapers.}
\label{fig:perfspec}

\end{figure}


Fig. \ref{fig:perfspec} shows the bias, variance and MSE variation of MT-MOGDF and MT-MAG with Thomson and multipeak tapers, with $N=8$. The MOGDF has a lesser MSE as compared to MT-MAG, even though its bias is larger. This establishes that the variance of spectral estimation takes a priority as opposed to the bias. Also, the variance and MSE values of MT-MOGDF with both tapers are better, and Thomson tapers give the least variance, across both phonemes. The multipeak tapers provide the best MSE for the phoneme /a/ as expected, since this was proposed to estimate peaked spectra. Similarly, Thomson tapers perform the best in terms of MSE with phoneme /s/.

\begin{figure}

\centering
\includegraphics[width=8cm, height=6cm]{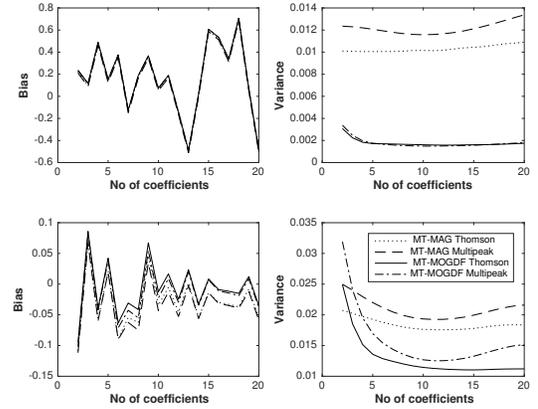}
\vspace{-0.4cm}
\caption{Cepstral-domain bias and variances for the phonemes /a/ (top row) and /s/ (bottom row). For MT-MOGDF, bias variation is similar to that of MT-MAG, while variance obtained is much lower compared to spectral-domain.}
\label{fig:perfceps}
\end{figure}

 The performance of both techniques in the cepstral-domain is considered, as given in Fig.~\ref{fig:perfceps}. Cepstrum representation using MT-MAG and MT-MOGDF is obtained by computing the DCT of the log spectrum as given in \eqref{eqn:ceps}. As expected, the bias and variance of both estimates are reduced as compared to the spectral-domain. The MT-MOGDF representation gives a lesser variance for nearly the same value of bias, compared to MT-MAG. The improvement of MT- MOGDF over MT-MAG is less significant in cepstral-domain, for the considered phonemes. Even here, MT-MOGDF with Thomson taper is found to have lesser variance. The performance trend in this case is similar to that seen in the spectral-domain.

\vspace{-0.3cm}
\section{Application: Speaker Recogniton}
\label{sec:spkrec}
 As an application-study of the proposed technique, a speaker recognition system is considered that uses the proposed MT-MOGDF features. The features chosen for comparison are the cepstral-domain features extracted using the periodogram, MOGDF [], MT-MAG and the MT-MOGDF spectrum estimates. The cepstral coefficients are then obtained by multiplying the estimated spectrum with the response of a Mel filter bank, followed by a logarithm operator. An associated GMM-UBM system is trained for a varying number of speakers and the corresponding EER and DCF (according to SRE10) values are tabulated in Tab.~\ref{tab:perf}. Among the considered methods, the MT-MOGDF features are most effective, and provide an improvement of around $20\%$ in comparison to the next-best MT-MAG feature set, averaged over the number of speakers. This improvement is attributed due to the reduction in the variance and high resolution properties of MT-MOGDF.

\begin{table}
\centering
\caption{DCF and EER performance of a speaker recognition system.}
\begin{tabular}{|P{0.85cm}|P{1cm}|P{0.85cm}|P{0.85cm}|P{0.85cm}|P{0.85cm}|}
\hline
Metric&No of Spk.&MFCC&MOGDF&MT-MAG&MT-MOGDF \\
\hline
{\begin{tabular} {c}DCF\\ $ \times 100$ \end{tabular}}&{\begin{tabular} {c}10\\20\\30\\40\\50\end{tabular}}&{\begin{tabular} {c}3.00   \\    4.50 \\       8.00\\        9.00 \\      9.00\end{tabular}}&{\begin{tabular} {c}0.30 \\   5.50 \\      6.00 \\      8.00 \\       8.00\end{tabular}}&{\begin{tabular} {c}1.00\\     3.50\\ 5.00 \\    6.86 \\    8.22\end{tabular} }&{\begin{tabular} {c}0.00  \\     2.50 \\     5.67 \\      6.00\\        6.40\end{tabular}}\\
\hline
EER&{\begin{tabular} {c}10\\20\\30\\40\\50\end{tabular}}&{\begin{tabular} {c}11.33\\10.00\\13.33\\12.50\\8.65\end{tabular}}&{\begin{tabular} {c}3.33\\5.00\\6.67\\6.98\\8.00\end{tabular}}&{\begin{tabular} {c}1.11\\5.00\\3.33\\5.00\\2.00\end{tabular}}&{\begin{tabular} {c}0.00\\1.31\\3.33\\2.50\\2.00\end{tabular}}\\
\hline
\end{tabular}
\label{tab:perf}
\end{table}

%

\vspace{-0.3cm}
\section{Conclusion}
\label{sec:conclu}
In this paper,  a novel representation for speech signals is proposed using multitaper-modified group delay functions. The proposed feature can be used in many speech applications, and in particular, among those that require information of speaker and speech. With different set of phonemes, it can be shown that the proposed method performs better that the existing multitaper-magnitude (MT-MAG) estimation technique, in terms of variance and MSE, both in spectral- and cepstral-domains. The performance of the proposed technique is in particular found to be the best with Thomson tapers. The computational requirements of the proposed technique is comparable to that of MT-MAG. Additionally, the utility of the proposed method is presented in a speaker recognition setup with an improvement of around $20\%$.

\balance

\bibliographystyle{IEEEtran}
\bibliography{letter_bib}
\end{document}